# A Hardware-Efficient Approach to Computing the Rotation Matrix from a Quaternion

Aleksandr Cariow[1] and Galina Cariowa[2]

[1] West Pomeranian University of Technology, Żołnierska 52, 71-210 Szczecin Poland
e-mail acariow@wi.zut.edu.pl
[2] West Pomeranian University of Technology, Żołnierska 52, 71-210 Szczecin Poland
e-mail gcariowa@wi.zut.edu.pl

**Abstract.** In this paper, we have proposed a novel VLSI-oriented approach to computing the rotation matrix entries from the quaternion coefficients. The advantage of this approach is the complete elimination of multiplications and replacing them by less costly squarings. Our approach uses Logan's identity, which proposes to replace the calculation of the product of two numbers on summing the squares via the Binomial Theorem. Replacing multiplications by squarings implies reducing power consumption as well as decreases hardware circuit complexity.

## 1 Introduction

Quaternions and rotation matrices are used especially in navigation, image encoding, machine vision, computer graphics, animation, and kinematics [1, 2]. Sometimes, in the above areas there is a need construction of the rotation matrix based on the vector described by quaternion. It is well known that 3×3 rotation matrix $\mathbf{R}_3$ can be expressed in term of quaternion

$$q = [q_0, q_1, q_2, q_3]^\mathrm{T}$$

as [2]:

$$\mathbf{R}_3 = \begin{bmatrix} q_0^2 + q_1^2 - q_2^2 - q_3^2 & 2(q_1q_2 - q_0q_3) & 2(q_0q_2 + q_1q_3) \\ 2(q_1q_2 + q_0q_3) & q_0^2 - q_1^2 + q_2^2 - q_3^2 & 2(q_2q_3 - q_0q_1) \\ 2(q_1q_3 - q_0q_2) & 2(q_0q_1 + q_2q_3) & q_0^2 - q_1^2 - q_2^2 + q_3^2 \end{bmatrix} \tag{1}$$

The above matrix is also called the direction cosine matrix. It is easy to see that the calculation of the elements of the rotation matrix requires 6 conventional multiplications, 4 operations of squaring, 6 trivial multiplications by 2 (shifts) and 15 additions.

There are a number of publications describing the rationalization of calculation of the rotation matrix entries [2, 3]. However, the solutions cited in these publications do not exhaust all possibilities of rationalization of computing. Below we will show how to get rid of having to perform multiplication operations with the help of replacing one nontrivial multiplication by one operation of squaring.

## 2 The approach

For a more compact representation, we introduce the following notation:

$$\mathbf{R}_3 = \begin{bmatrix} c_{0,0} & c_{0,1} & c_{0,2} \\ c_{1,0} & c_{1,1} & c_{1,2} \\ c_{2,0} & c_{2,1} & c_{2,2} \end{bmatrix} \tag{2}$$

where

$$c_{0,0} = q_0^2 + q_1^2 - q_2^2 - q_3^2, \, c_{0,1} = 2(q_1 q_2 - q_0 q_3), \; c_{0,2} = 2(q_0 q_2 + q_1 q_3),$$
$$c_{1,0} = 2(q_1 q_2 + q_0 q_3), \; c_{1,1} = q_0^2 - q_1^2 + q_2^2 - q_3^2, \; c_{1,2} = 2(q_2 q_3 - q_0 q_1),$$
$$c_{2,0} = 2(q_1 q_3 + q_0 q_2), \; c_{2,1} = 2(q_0 q_1 - q_2 q_3), \; c_{2,2} = q_0^2 - q_1^2 - q_2^2 + q_3^2.$$

In 1971, Logan noted that the multiplication of two numbers can be performed using the following expression [4, 5]:

$$ab = \frac{1}{2}[(a+b)^2 - a^2 - b^2] \tag{3}$$

Using the Logan's identity we can write:

$$2(q_1 q_2 - q_0 q_3) = (q_1 + q_2)^2 - (q_0 + q_3)^2 - (q_1^2 + q_2^2) + (q_0^2 + q_3^2),$$
$$2(q_2 q_3 + q_0 q_1) = (q_2 + q_3)^2 + (q_0 + q_1)^2 - (q_1^2 + q_2^2) - (q_0^2 + q_3^2),$$
$$2(q_0 q_3 + q_1 q_2) = (q_0 + q_3)^2 + (q_1 + q_2)^2 - (q_1^2 + q_2^2) - (q_0^2 + q_3^2),$$
$$2(q_0 q_3 - q_0 q_2) = (q_0 + q_3)^2 - (q_0 + q_2)^2 + (q_0^2 + q_2^2) - (q_0^2 + q_3^2),$$
$$2(q_1 q_3 + q_0 q_2) = (q_1 + q_3)^2 + (q_0 + q_2)^2 - (q_1^2 + q_2^2) - (q_0^2 + q_3^2),$$
$$2(q_2 q_3 - q_0 q_1) = (q_2 + q_3)^2 - (q_0 + q_1)^2 + (q_1^2 - q_2^2) + (q_0^2 - q_3^2).$$

Then the all rotation matrix entries, that previously required performing the multiplications, can be calculated only with the help of squaring operations:

$$c_{0,1} = (q_1 + q_2)^2 - (q_0 + q_3)^2 - (q_1^2 + q_2^2) + (q_0^2 + q_3^2),$$
$$c_{0,2} = (q_2 + q_3)^2 + (q_0 + q_1)^2 - (q_1^2 + q_2^2) - (q_0^2 + q_3^2),$$

$$c_{1,0} = (q_0 + q_3)^2 + (q_1 + q_2)^2 - (q_1^2 + q_2^2) - (q_0^2 + q_3^2)\,,$$
$$c_{1,2} = (q_0 + q_3)^2 - (q_0 + q_2)^2 + (q_0^2 + q_2^2) - (q_0^2 + q_3^2)\,,$$
$$c_{2,0} = (q_1 + q_3)^2 + (q_0 + q_2)^2 - (q_1^2 + q_2^2) - (q_0^2 + q_3^2)\,,$$
$$c_{2,1} = (q_2 + q_3)^2 - (q_0 + q_1)^2 + (q_1^2 - q_2^2) + (q_0^2 - q_3^2)\,.$$

But, by observing that the expressions above contain duplicate sums, one can easily reduce the total number of additions.

Introduce the following notations:

$$(q_1 + q_2)^2 = \phi_0\,,\ (q_0 + q_3)^2 = \phi_1\,,\ (q_2 + q_3)^2 = \phi_2\,,\ (q_0 + q_1)^2 = \phi_3\,,$$
$$(q_1 + q_3)^2 = \phi_4\,,\ (q_0 + q_2)^2 = \phi_5\,,\ q_1^2 + q_2^2 = \vartheta_0\,,\ q_0^2 + q_3^2 = \vartheta_1\,,$$
$$q_0^2 + q_2^2 = \vartheta_2\,,\ q_0^2 - q_2^2 = \vartheta_3\ q_0^2 - q_3^2 = \vartheta_4\,,\ \vartheta_0 + \vartheta_1 = \lambda\,. \tag{4}$$

Then we can finally write:

$$c_{0,0} = \vartheta_3 + \vartheta_4\ \ c_{0,1} = (\phi_0 - \phi_1) - (\vartheta_0 - \vartheta_1)\,,\ \ c_{0,2} = (\phi_2 + \phi_3) - \lambda\,,$$
$$c_{1,0} = (\phi_0 + \phi_1) - \lambda\,,\ \ c_{1,2} = (\phi_1 - \phi_5) + (\vartheta_2 - \vartheta_1)\,,\ c_{1,1} = \vartheta_4 - \vartheta_3\,,$$
$$c_{2,0} = (\phi_4 + \phi_5) - \lambda\,,\ \ c_{2,1} = (\phi_2 - \phi_3) + (\vartheta_3 - \vartheta_4)\,,\ c_{2,2} = \vartheta_1 - \vartheta_0\,. \tag{5}$$

It is easily seen that the calculation of the above expressions requires 10 squarings and 29 additions.

Fig. 1 shows the overall structure of the processing unit for calculating the entries of the rotation matrix using expressions (5) and Figs 2-5 show the structures of the blocks that make up the unit.

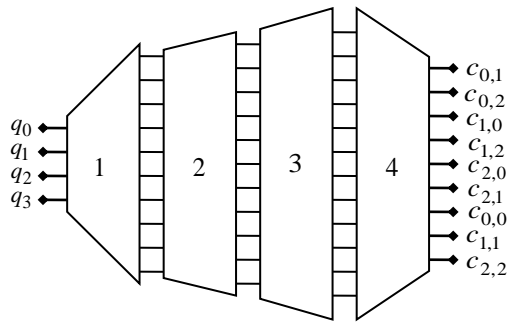

**Fig. 1.** Overall structure of the processing unit for calculating the entries of the rotation matrix in correspondence to the proposed approach.

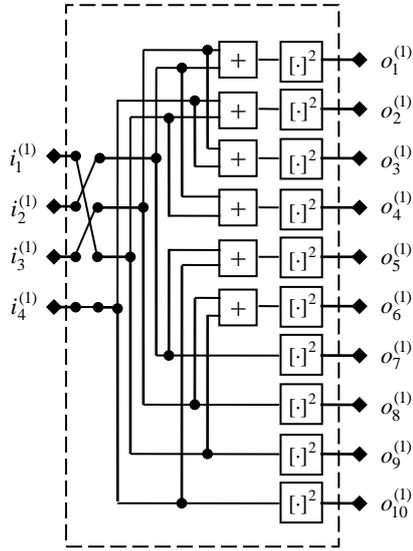

**Fig. 4.** *The structure of 1ˢᵗ block.*

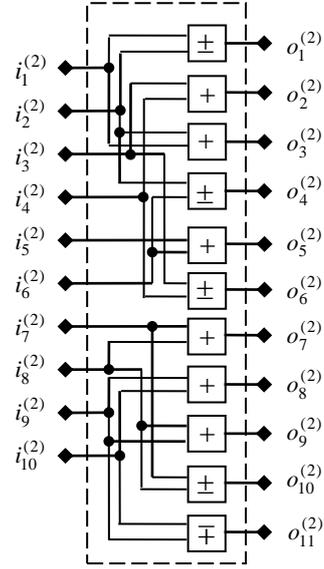

**Fig. 3** *The structure of 2ⁿᵈ block.*

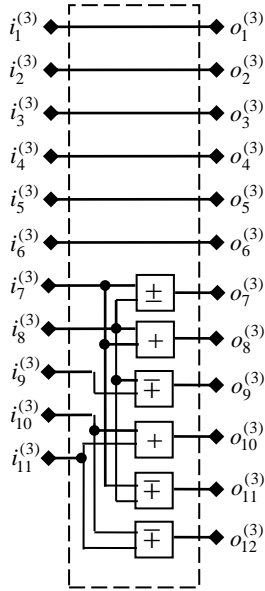

**Fig. 4.** *The structure of 3ʳᵈ block.*

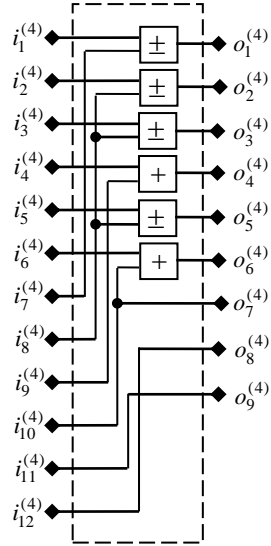

**Fig. 5.** *The structure of 4ᵗʰ block.*

## 3       Conclusion

It should be noted that squares are a special case of multiplication where both operands are identical. For this reason designers often use general-purpose multipliers to implement the squaring units by connecting a multiplier's inputs together. Even though using general-purpose multipliers that are available as part of design packages reduces design time, it results in increased area and power requirements for the design [6]. Meanwhile, since the two operands are identical, some rationalizations can be made during the implementation of a dedicated squarer. In particular, unlike the general-purpose multiplier a dedicated squaring unit will have only one input, which allows to simplify the circuit. The article [7] shows that the dedicated squaring unit requires less than half whole amount of the logic gates as compared to the general-purpose multiplier. Dedicated squarer is area efficient, consumes less energy and dissipates less power as compared to general purpose multiplier. So, it is easy to estimate that our approach is more efficient in terms of the discussed parameters than the direct calculation of the rotation matrix entries in accordance with (1).

## References

1. Markley, F.L.: 'Unit Quaternion from Rotation Matrix', *Journal of Guidance, Control, and Dynamics*, vol. 31, No. 2, pp. 440-442, 2008, doi: 10.2514/1.31730.
2. Shuster M.D., Natanson G.A.: 'Quaternion Computation from a Geometric Point of View', *The Journal of Astronautical Sciences*, vol. 41, no. 4, pp. 545–556, 1993.
3. Doukhnitch, E., Chefranov A., Mahmoud A. - 'Encryption Schemes with Hyper-Complex Number Systems and Their Hardware-Oriented Implementation' in Elci, A. (Ed): 'Theory and Practice of Cryptography Solutions for Secure Information Systems' (Hershey, Pa.: IGI Global 2013).
4. Logan, J.R.: 'A square-summing high-speed multiplier', Comput. Des., 1971, pp. 67-70.
5. Johnson, E.L.: 'A Digital Quarter Square Multiplier', IEEE Transactions on Computers, vol. C-29, no. 3, pp. 258 − 261, 1980, doi:10.1109/TC.1980.1675558.
6. Deshpande, A., Draper, J.: 'Squaring units and a comparison with multipliers'. 53rd IEEE International Midwest Symposium on Circuits and Systems (MWSCAS 2010), Seattle, Washington August 1st-4th, 2010, pp. 1266 − 1269, doi: 10.1109/MWSCAS.2010.5548763.
7. Liddicoat A. A., Flynn M. J.: 'Parallel Computation of the Square and Cube Function', Computer Systems Laboratory; Stanford University, Technical report No. CSL-TR-00-808, August 2000.